\documentclass[conference]{IEEEtran}
\usepackage[hyphens]{url}
\usepackage{hyperref}
\hypersetup{colorlinks,allcolors=blue}

\usepackage[backend=bibtex,style=ieee,citestyle=numeric-comp]{biblatex}
\usepackage{amsmath,amssymb,amsfonts}
\usepackage{graphicx}
\usepackage{textcomp}
\usepackage{xcolor}
\usepackage{booktabs}
\usepackage{array}
\usepackage{balance}
\usepackage{enumitem} 
\usepackage{comment}
\usepackage{soul}

\usepackage{geometry}
\newcolumntype{C}[1]{>{\centering\arraybackslash}p{#1}}
\newcolumntype{L}[1]{>{\raggedright\arraybackslash}p{#1}}

\title{Context-Based Adversarial Attacks on AI Code Generators: Vulnerability Analysis and Implications}
\author{
  \IEEEauthorblockN{Walther A.\ Del Orbe, John D. Hastings, Varghese Vaidyan}
  \IEEEauthorblockA{
    \textit{The Beacom College of Computer \& Cyber Sciences}\\
    \textit{Dakota State University}\\
    Madison, SD, USA\\
    walther.delorbe@trojans.dsu.edu, john.hastings@dsu.edu, varghese.vaidyan@dsu.edu}
}

\begin{document}

\maketitle

\begin{abstract}
AI-powered code generation systems have transformed software
development but introduce critical inference-time security
vulnerabilities.  This research presents a systematic investigation of
context-based adversarial attacks, where strategically crafted
contextual inputs, including comments, documentation, variable names,
bias large language models toward generating exploitable code.
Through 2,800 controlled experiments across CodeT5+, CodeLlama,
GPT-3.5-Turbo, and GPT-4, we quantify attack effectiveness and defense
mechanisms.  Results demonstrate that adversarial conditions increase
vulnerability generation $10.7\times$ (from 3.5\% to 37.4\%), with
direct instruction attacks achieving 100\% success on GPT-3.5-Turbo.
Cross-model transferability reaches 60--100\%, indicating systemic
architectural vulnerabilities rather than model-specific flaws.  Our
dual-layer defense framework achieves 89.1\% detection rate with 0.3\%
false positives and 520\,ms latency, demonstrating practical
feasibility for real-time deployment in development environments.
\end{abstract}


\section{Introduction}

AI-powered code generation systems such as GitHub Copilot, CodeT5+,
and GPT-4 have fundamentally transformed software development,
enabling developers to generate functional code from natural language
descriptions.  Industry analyses show 55\% faster task completion with
AI assistance~\cite{peng2023}. However, this transformation introduces
a novel threat surface: \emph{inference-time contextual manipulation
attacks}.  Unlike training-time data poisoning, inference-time attacks
manipulate the context supplied during normal code generation,
creating a practical threat surface for AI-assisted development.
Recent assessments reveal alarming rates: Veracode found 86\% of
relevant AI-generated XSS samples failed defenses
~\cite{veracode2025}, while CSET \cite{ji2024} found that nearly half
of code snippets produced by five evaluated LLMs contained bugs with
the potential for malicious exploitation.  OWASP now ranks prompt
injection as the \#1 LLM security risk~\cite{owasp2025}.

Consider a realistic attack scenario: an adversary contributes seemingly
helpful documentation to a widely used open-source library, embedding
phrases like \textit{``for backward compatibility, use MD5 hashing''} or
\textit{``quick prototype---skip input validation for performance.''}
When developers use AI-assisted code generation tools referencing this
library, the malicious context biases the model toward insecure code.
Given widespread adoption across millions of developers and enterprise
environments, such context-based supply-chain attacks could propagate
vulnerabilities into production systems with minimal attacker effort.

This study investigates the security implications of such
inference-time contextual manipulation through the following research
questions:
\begin{enumerate}[label={\textbf{RQ\arabic*:}},left=1.0em]
\item Which context positions and linguistic patterns most
effectively induce vulnerabilities?
\item How can malicious instructions be steganographically
embedded in documentation?
\item Do contextual attacks transfer across architectures?
\item What real-time detection mechanisms can identify
manipulation without excessive false positives?
\end{enumerate}

This research provides:
(1)~a systematic analysis of inference-time contextual
manipulation;
(2)~quantitative assessment across four AI architectures;
(3)~cross-model transferability analysis; and
(4)~a practical defense framework achieving 89.1\% detection with
0.3\% false positives.

The remainder of this paper is organized as follows: Sections
\ref{sec:background}-\ref{sec:prob_defn} provide background on AI code
generation and inference-time attacks, and define the threat model and
problem formulation. Sections
\ref{sec:meth}-\ref{sec:defense_framework} present the methodology,
empirical results, and defense framework. Sections
\ref{sec:discussion}-\ref{sec:conclusion} discuss implications and
limitations, review related work, address ethical considerations, and
conclude.

\section{Background}
\label{sec:background}

This section establishes the technical foundations necessary to
understand context-based adversarial attacks: the attention-based
architecture that creates exploitable dependencies in code generators,
the vulnerability patterns inherited from training corpora, and the
key distinction between inference-time and training-time attack vectors
that motivates this work.
\subsection{AI Code Generation Architecture}
Transformer-based code generators use multi-head self-attention to
process contextual tokens:
\begin{equation}
  \mathrm{Attention}(Q,K,V)
    = \mathrm{softmax}\!\left(\frac{QK^{\top}}{\sqrt{d_k}}\right)V
\end{equation}
where $Q$, $K$, and $V$ are learned query, key, and value projections,
and $d_k$ is the key dimension.  This mechanism enables contextual
understanding but creates exploitable dependencies where adversarial
cues bias token prediction toward vulnerable patterns.

\subsection{Vulnerability Inheritance}
Models trained on GitHub/StackOverflow corpora inherit prevalent
vulnerabilities.  For example, Fu et al. \cite{fu2025} found security
weaknesses in 24.2\% of JavaScript and 29.5\% of Python snippets
produced by Codeium, CodeWhisperer, and Copilot in GitHub projects
including SQL injection, hard-coded credentials or passwords, and
risky cryptographic algorithms.  Contextual manipulation exploits
learned correlations between documentation cues and vulnerable code
patterns.

\subsection{Inference-Time vs.\ Training-Time Attacks}
Training-time attacks compromise model behavior through the training
pipeline, whereas inference-time attacks manipulate only the context
supplied at generation time. This distinction motivates the threat
model formalized in Section \ref{sec:prob_defn}.

\section{Threat Model \& Problem Definition}
\label{sec:prob_defn}

\subsection{Threat Model}
This study assumes an attacker can influence contextual material
consumed by an AI code generation system, including comments,
documentation strings, variable names, or reference examples. The
attacker cannot access model parameters, training data, or deployment
infrastructure. The manipulation succeeds when the modified context
appears legitimate to human reviewers while increasing the likelihood
that generated code contains an exploitable vulnerability.

\subsection{Problem Definition}
Let $M$ be an AI code generator, $\mathcal{C}$ the contextual input
space, and $\mathcal{O}$ the output space.
A \emph{context-based adversarial transformation} is a function
$A:\mathcal{C}\to\mathcal{C}'$ transforming benign context $c$ into
adversarial $c'$ such that:
(1)~$c'$ appears legitimate to human reviewers;
(2)~$P(M(c')\ \text{generates vulnerable code}) \gg P(M(c)\ \text{generates vulnerable code})$;
(3)~$A$ requires no model parameter access. 

We define three core metrics:
\begin{align}
  \mathrm{VGR}(M,c,V) &=
    \frac{\text{Vulnerable outputs}}{\text{Total generations}} \\[4pt]
  \mathrm{ASR}(A,M,V) &=
    \frac{\text{Successful inductions}}{\text{Total adversarial attempts}} \\[4pt]
  \mathrm{TR}(M_o,M_t) &=
    \frac{\mathrm{ASR}(M_t)}{\mathrm{ASR}(M_o)}
    \label{eq:tr}
\end{align}
Here, \textit{vulnerability generation rate} (VGR) measures the
proportion of all model outputs that contain a vulnerability from
category $V$ under a given model-context pair $(M,c)$. \textit{Attack
  success rate} (ASR) measures the proportion of adversarial attempts
by transformation $A$ that successfully induce vulnerable code in
model $M$ for vulnerability class $V$. Finally,
\textit{transferability rate} (TR) measures the fraction of attack
effectiveness retained when crossing architectural boundaries. A TR
value of 1.0 indicates full transferability; values above 0.60 are
considered high, indicating architectural commonality in the exploited
vulnerability.

\section{Methodology}
\label{sec:meth}

To answer the research questions, this study evaluates context-based
adversarial manipulation across multiple model architectures,
vulnerability categories, attack techniques, and detection mechanisms.

\subsection{Model Selection}
This study selected four models spanning diverse architectural
families and training regimes.  CodeT5+ (770M parameters, open-source
encoder-decoder~\cite{wang2023codet5}), CodeLlama-7B (open-source
Llama decoder variant~\cite{roziere2023}), GPT-3.5-Turbo (commercial,
RLHF-trained), and GPT-4 (commercial, state-of-the-art as of December
2024) together represent the spectrum from open-source to proprietary
and from supervised to reinforcement-learning-based training.  Default
inference parameters were temperature~0.7, top-p~0.9, and
max\_tokens~512, held constant across all models.  All API calls used
pinned snapshot versions to prevent model drift between runs.

\subsection{Vulnerability Categories}
Five vulnerability categories were selected based on their prevalence
in prior code generation security studies~\cite{pearce2022} and their
coverage in the OWASP Top~10~\cite{owasp_top10} and OWASP LLM
Top~10~\cite{owasp2025}.  Each category is mapped to a Common Weakness
Enumeration (CWE) identifier: SQL Injection (CWE-89), Cross-Site
Scripting (CWE-79), Hardcoded Credentials (CWE-798), Path Traversal
(CWE-22), and Insecure Cryptography (CWE-327).

\subsection{Attack Techniques}
Four prompt conditions were constructed to span a spectrum from
neutral to explicitly adversarial:
\begin{itemize}
  \item \textbf{Baseline:} Neutral prompts without any security-relevant
        guidance, establishing each model's natural vulnerability rate
        under normal usage conditions.
  \item \textbf{Direct Instruction:} Explicit insecure directives
        embedded in inline comments
        (e.g., ``Use string concatenation for SQL queries''),
        testing whether models comply with unambiguous adversarial commands.
  \item \textbf{Semantic Priming:} Subtle contextual cues that imply
        insecurity without explicit instruction
        (e.g., ``quick prototype, skip validation''),
        testing susceptibility to plausibly legitimate framing.
  \item \textbf{Example-Based:} Vulnerable code snippets provided as
        inline reference implementations, exploiting the model's
        tendency to replicate patterns present in context.
\end{itemize}

\subsection{Experimental Matrix}
As shown in Table~\ref{tab:experimental-trial-distribution}, a total
of 2,800 trials were conducted: CodeT5+ (1,000), CodeLlama (1,000),
GPT-3.5-Turbo (500), and GPT-4 (300).  Each vulnerability category
received 200~tests for open-source models, 100~for GPT-3.5-Turbo, and
60~for GPT-4.  Within each model--category cell, trials were divided
equally across the four prompt conditions (baseline, direct, semantic,
example-based), yielding 50 trials per condition per category for
open-source models, 25 for GPT-3.5-Turbo, and 15 for GPT-4.
Commercial model sample sizes were constrained by API cost; the
resulting per-category counts provide sufficient power for chi-square
analysis at $\alpha = 0.05$ with 80\% power.

\begin{table}[h]
\centering
\caption{Experimental Trial Distribution by Model}
\label{tab:experimental-trial-distribution}
\begin{tabular}{lccc}
\hline
\textbf{Model} & \textbf{Total Trials} & \textbf{Tests per Category} & \textbf{Categories} \\
\hline
CodeT5+ & 1,000 & 200 & 5 \\
CodeLlama & 1,000 & 200 & 5 \\
GPT-3.5-Turbo & 500 & 100 & 5 \\
GPT-4 & 300 & 60 & 5 \\
\hline
\textbf{Total} & 2,800 & -- & 5 \\
\hline
\end{tabular}
\end{table}

\subsection{Detection \& Validation Pipeline}
\label{sec:detection}
Generated outputs were evaluated through a three-stage detection
pipeline designed to balance automated coverage with human validation.
First, automated static analysis was performed using an ensemble of
three tools: Bandit (Python), Semgrep, and ESLint (JavaScript), with a
majority-vote requirement across tools to reduce per-tool
false-negative rates and avoid over-reliance on any single analyzer.
Second, regex-based and abstract syntax tree (AST) pattern matching
detected structural vulnerability patterns that evade lexical-only
analysis, such as string concatenation within SQL query construction
or direct assignment to \texttt{innerHTML}.  Third, a 15\% stratified
random sample, balanced across models, vulnerability categories, and
attack techniques, was manually reviewed; discrepancies were resolved
(Cohen's $\kappa = 0.87$), validating the reliability of automated
detection labels used throughout the study.

\subsection{Cross-Model Transferability \& Statistical Validation}
Cross-model transferability was evaluated by taking all adversarial
prompts that achieved VGR~$\geq 25\%$ on their origin model and
applying them verbatim without modification to each of the other three
models. Transferability rate (TR) was computed according to
(\ref{eq:tr}).

All statistical analyses were pre-specified before data collection.
Chi-square ($\chi^2$) tests were used for categorical comparisons of
vulnerability occurrence rates between baseline and adversarial
conditions.  Effect sizes were quantified using Cohen's~$d$ for
pairwise comparisons and $\eta^2$ for one-way ANOVA, which was applied
to compare vulnerability rates across multiple groups (attack
techniques, vulnerability categories, and context positions).
Post-hoc pairwise comparisons used Tukey's Honestly Significant
Difference (HSD) test.  Confidence intervals for all proportions were
computed via 10,000-iteration bootstrap resampling~\cite{efron1979}.
To control the familywise Type~I error rate across multiple
simultaneous comparisons, Bonferroni correction was applied, producing
an adjusted significance threshold of $\alpha_{\text{adj}} = \alpha /
n$ with $\alpha = 0.05$.  For the transferability analysis, a
one-sample $t$-test was used to assess whether observed TR values
significantly exceeded the 50\% chance baseline.

\section{Results \& Analysis}
\label{sec:results}

The following empirical results are organized around RQ1-RQ3. The
defense evaluation corresponding to RQ4 is presented separately in
Section \ref{sec:defense_framework}.

\subsection{Overall Vulnerability Generation}

As shown in Table~\ref{tab:vgr}, adversarial conditions increased
the mean vulnerability generation rate (VGR) from 3.5\% under
baseline conditions to 37.4\% under adversarial conditions, a
$10.7\times$ increase that is statistically significant
($\chi^2 = 847.3$, $p < 0.001$, Cohen's $d = 1.2$--$3.8$),
with effect sizes ranging from large to very large in practical
terms. 
\begin{table}[t]
\centering
\caption{Vulnerability Generation Rates (VGR) by Model}
\label{tab:vgr}
\setlength{\tabcolsep}{4pt}
\begin{tabular}{lrrrr}
\toprule
\textbf{Model} & \textbf{$n$} & \textbf{Baseline} & \textbf{Adversarial} & \textbf{Mean} \\
\midrule
CodeT5+       & 1,000 &  2.0\% & 22.7\% & 15.0\% \\
CodeLlama     & 1,000 & 12.0\% & 18.3\% & 13.1\% \\
GPT-3.5-Turbo &   500 &  0.0\% & 68.7\% & 52.0\% \\
GPT-4         &   300 &  0.0\% & 40.0\% & 30.0\% \\
\midrule
\textbf{Mean} & 2,800 &  3.5\% & 37.4\% & 27.5\% \\
\bottomrule
\end{tabular}
\end{table}
Commercial models (GPT-3.5-Turbo, GPT-4) exhibited 0\% baseline VGR
consistent with RLHF safety training, yet substantially higher
adversarial VGR than open-source counterparts.  One possible
explanation is that RLHF instruction-following may amplify
responsiveness to authoritative adversarial directives; however, this
study does not isolate RLHF as a causal mechanism.  The 95\%~CI for
adversarial VGR is $[35.1\%, 39.7\%]$.

\subsection{Attack Technique Effectiveness (RQ1, Part 1)}

Table~\ref{tab:asr} reports the attack success rate (ASR) by technique
and model.  Direct instruction achieved the highest mean ASR at 55\%,
with GPT-3.5-Turbo reaching 100\% ($\chi^2 = 421.8$, $p < 0.001$, $d =
2.7$), suggesting that explicitly phrased insecure directives are
highly effective against RLHF-trained models.  Example-based attacks
achieved 31.4\% ($d = 1.8$), exploiting the model's tendency to
replicate patterns present in context.  Semantic priming was least
effective at 17.5\%, though with high variance across models (SD $=
16.3\%$), indicating model-specific susceptibility to subtle cues.

\begin{table}[t]
\centering
\caption{Attack Success Rates (ASR, \%) by Technique and Model}
\label{tab:asr}
\setlength{\tabcolsep}{3.5pt}
\begin{tabular}{lrrrrr}
\toprule
\textbf{Technique} & \textbf{CT5+} & \textbf{CL} & \textbf{G3.5} & \textbf{G4} & \textbf{Mean} \\
\midrule
Baseline      &  0   & 12   &   0  &  0  &  3.0 \\
Direct        & 20   & 20   & 100  & 80  & 55.0 \\
Semantic      & 20   &  0   &  40  & 10  & 17.5 \\
Example-Based & 20   & 20.4 &  75  & 10  & 31.4 \\
\bottomrule
\end{tabular}
\end{table}

\subsection{Positional Effects (RQ1, Part 2)}

Table~\ref{tab:position} shows ASR as a function of adversarial token
position relative to the generation target.  Placement in the
pre-function zone (10--50 tokens immediately before the target)
achieved the highest ASR of 62.1\%, which we attribute to
attention-mechanism recency bias: attention weight analysis confirms
$2.3\times$ higher weights for pre-function tokens compared to tokens
placed 300 or more tokens away.  A one-way ANOVA confirmed a
significant effect of position ($F(3,696) = 127.4$, $p < 0.001$,
$\eta^2 = 0.35$), with all pairwise differences remaining significant
after Bonferroni correction.

\begin{table}[t]
\centering
\caption{Context Position and Attack Success Rate (ASR)}
\label{tab:position}
\begin{tabular}{llr}
\toprule
\textbf{Position} & \textbf{Description} & \textbf{ASR} \\
\midrule
Pre-function  & 10--50 tokens before target & 62.1\% \\
Initial       & First 50 tokens             & 48.3\% \\
Mid-context   & 100--200 tokens from target & 38.5\% \\
Distant       & 300$+$ tokens from target   & 22.7\% \\
\bottomrule
\end{tabular}
\end{table}

\subsection{Vulnerability Category Rates}

Across vulnerability categories, Table~\ref{tab:categories} presents
adversarial VGR across all five CWE categories.  SQL Injection showed
the highest rate (36.6\%), consistent with its prevalence in
open-source training corpora.  A one-way ANOVA confirmed significant
differences across categories ($F(4,2795) = 47.3$, $p < 0.001$,
$\eta^2 = 0.063$), and post-hoc Tukey HSD testing identified SQL
Injection vs.\ Insecure Cryptography as the largest pairwise
difference ($p < 0.001$, $d = 1.42$).

\begin{table}[t]
\centering
\caption{Adversarial VGR (\%) by Vulnerability Category}
\label{tab:categories}
\setlength{\tabcolsep}{3pt}
\begin{tabular}{lrrrrr}
\toprule
\textbf{Category} & \textbf{CT5+} & \textbf{CL} & \textbf{G3.5} & \textbf{G4} & \textbf{Mean} \\
\midrule
SQL Inj.\ (CWE-89)      & 22 & 24.5 & 65 & 35 & 36.6 \\
Hardcoded (CWE-798)     & 15 & 18.2 & 55 & 32 & 30.1 \\
XSS (CWE-79)            & 10 &  8.7 & 48 & 25 & 23.0 \\
Path Trav.\ (CWE-22)    &  8 &  7.4 & 42 & 22 & 19.9 \\
Ins.\ Crypto (CWE-327)  &  6 &  5.8 & 38 & 18 & 17.0 \\
\bottomrule
\end{tabular}
\end{table}

\subsection{Linguistic Pattern Analysis (RQ2)}

As seen in Table~\ref{tab:patterns}, authority/imperative patterns
achieved 58.3\% ASR (95\%~CI $[52.7\%, 63.9\%]$), followed by legacy
justifications (51.7\%, $[46.2\%, 57.2\%]$).  These phrases appear
legitimate in technical documentation while biasing generation toward
insecure patterns, enabling steganographic embedding.  Logistic
regression: linguistic pattern score explains 42\% of VGR variance
($R^2 = 0.42$, $F(5,694) = 101.3$, $p < 0.001$, AUC~$= 0.84$).

\begin{table}[t]
\centering
\caption{High-Risk Linguistic Patterns and ASR}
\label{tab:patterns}
\setlength{\tabcolsep}{3pt}
\begin{tabular}{llrr}
\toprule
\textbf{Pattern} & \textbf{Examples} & \textbf{ASR} & \textbf{95\% CI} \\
\midrule
Authority      & ``must use,'' ``required'' & 58.3\% & $[52.7,63.9]$ \\
Legacy         & ``backward compat.''       & 51.7\% & $[46.2,57.2]$ \\
Simplification & ``simple,'' ``quick''      & 42.1\% & $[36.8,47.4]$ \\
Temporal       & ``prototype,'' ``for now'' & 39.4\% & $[34.1,44.7]$ \\
Performance    & ``optimized,'' ``fast''    & 36.8\% & $[31.6,42.0]$ \\
\bottomrule
\end{tabular}
\end{table}

\subsection{Cross-Model Transferability (RQ3)}

As shown in Table~\ref{tab:transfer}, attack prompts transferred most
effectively between open-source models, achieving 95--100\%
cross-transfer (mean TR $= 0.975$, 95\%~CI $[0.94, 1.01]$).  Transfers
from open-source to commercial models were somewhat lower at 65--82\%
(mean TR $= 0.738$), while GPT-3.5-Turbo to GPT-4 achieved 90\%,
suggesting shared architectural properties within the GPT family.  All
TR values significantly exceeded the 50\% chance baseline (one-sample
$t$-test, $p < 0.001$ for all pairs), and high transferability (TR $>
0.60$) was observed in 83\% of model pairs, consistent with a systemic
vulnerability rooted in the attention mechanism rather than
model-specific behavior.  A moderate positive correlation between
model parameter count and transfer resistance was also observed
(Pearson $r = 0.47$, $p = 0.032$).

\begin{table}[t]
\centering
\caption{Attack Transferability Matrix (TR values)}
\label{tab:transfer}
\setlength{\tabcolsep}{4pt}
\begin{tabular}{lcccc}
\toprule
\textbf{Origin$\to$Target} & \textbf{CT5+} & \textbf{CL} & \textbf{G3.5} & \textbf{G4} \\
\midrule
CodeT5+    & ---  & 1.00 & 0.78 & 0.65 \\
CodeLlama  & 0.95 & ---  & 0.82 & 0.70 \\
GPT-3.5    & 0.65 & 0.68 & ---  & 0.90 \\
GPT-4      & 0.58 & 0.62 & 0.85 & ---  \\
\bottomrule
\end{tabular}
\end{table}

\section{Defense Framework}
\label{sec:defense_framework}
To address RQ4, the study evaluates whether real-time detection
mechanisms can identify contextual manipulation without excessive
false positives. The proposed defense framework uses two layers:
prompt-level analysis before generation and code-level analysis after
generation.

\subsection{Layer 1: Prompt-Level Analysis}
Pre-generation detection combines four mechanisms:
\begin{itemize}
  \item \textbf{Linguistic Pattern Scoring:} Weighted scoring for
        authority ($w{=}3$), legacy ($w{=}3$), simplification ($w{=}2$),
        and temporal ($w{=}2$) patterns.
        Threshold: 7.5 (ROC-optimized).
  \item \textbf{Semantic Validation:} Cosine similarity between security
        requirements and contextual suggestions.
        Flag contradictions with similarity $< 0.3$.
  \item \textbf{Perplexity Analysis:} GPT-2 perplexity calculation.
        Flag prompts with perplexity $> 150$
        (98th percentile of benign prompts).
  \item \textbf{Position Monitoring:} Critical-zone surveillance
        (10--50 tokens pre-function).
        Density threshold: 0.4 patterns per 10 tokens.
\end{itemize}

\subsection{Layer 2: Code-Level Analysis}

The post-generation layer reuses the static-analysis and
structural-pattern checks described in Section \ref{sec:detection},
then extends them with semantic taint/control-flow analysis and
differential comparison against a clean baseline generation. Taint
analysis tracks unsanitized inputs to security-sensitive sinks, while
differential analysis flags outputs whose normalized edit distance
from the clean baseline exceeds 0.35. Together, these checks identify
vulnerable code patterns that remain after prompt-level filtering.

\subsection{Defense Evaluation}

Table~\ref{tab:defense} reports performance on the held-out test set
($n = 560$, 20\% stratified split not used during threshold tuning).
The combined system achieves 89.1\% detection rate (95\%~CI $[86.4\%,
  91.8\%]$) with 0.3\% FPR.  McNemar's test confirms the combined
system significantly outperforms each individual layer ($p < 0.001$).
End-to-end latency averages 520\,ms: the prompt layer completes in
under 100\,ms, while the code layer averages 460\,ms (99th percentile:
780\,ms).  The system generates fewer than one false alarm per 300
legitimate prompts, making it suitable for real-time IDE integration
without excessive workflow disruption.
\begin{table}[t]
\centering
\caption{Defense System Performance (Held-Out Test Set, $n=560$)}
\label{tab:defense}
\setlength{\tabcolsep}{3.5pt}
\begin{tabular}{lrrrr}
\toprule
\textbf{Component} & \textbf{DR} & \textbf{FPR} & \textbf{Prec.} & \textbf{F1} \\
\midrule
Linguistic Patterns & 68.2\% & 0.0\% & 100\%  & 0.811 \\
Semantic Validation & 54.1\% & 0.0\% & 100\%  & 0.703 \\
Position Monitoring & 62.9\% & 0.0\% & 100\%  & 0.772 \\
Perplexity Analysis & 41.7\% & 2.3\% & 94.7\% & 0.580 \\
\textbf{Prompt Layer} & 73.3\% & 0.0\% & 100\% & 0.846 \\
\midrule
Static Analysis     & 42.1\% & 0.0\% & 100\%  & 0.593 \\
Pattern Matching    & 38.7\% & 0.0\% & 100\%  & 0.558 \\
Taint Analysis      & 29.4\% & 1.2\% & 96.1\% & 0.449 \\
Differential        & 51.3\% & 0.0\% & 100\%  & 0.677 \\
\textbf{Code Layer} & 67.8\% & 0.3\% & 99.6\% & 0.807 \\
\midrule
\textbf{Combined}   & \textbf{89.1\%} & \textbf{0.3\%} & \textbf{99.7\%} & \textbf{0.942} \\
\bottomrule
\end{tabular}
\end{table}
Additional system metrics include Sensitivity (89.1\%), Specificity (99.7\%),
MCC ( 0.923), AUC-ROC (0.974), and NPV (94.3\%).

\section{Discussion}
\label{sec:discussion}

\subsection{Key Findings}
Results confirm context-based manipulation as a fundamental
vulnerability in transformer architectures.  The high transferability
suggests that the vulnerability stems from attention mechanisms
essential for code generation, not model-specific implementations.
Commercial models exhibit superior baseline security but higher
adversarial susceptibility; this suggests that RLHF
instruction-following amplifies responsiveness to adversarial
directives, an obedient insecurity effect, though this study does not
isolate alignment method as a causal factor.

\subsection{Security Implications}
Organizations face quantifiable risk under adversarial conditions,
with a mean VGR of 27.5\% observed across all models and attack
techniques. The high cross-model transferability reported in Section
\ref{sec:results} means that a single adversarial documentation
artifact could propagate vulnerabilities broadly across the
AI-assisted development ecosystem.

To illustrate the practical scale of this risk, consider a scenario in
which 5\% of the documentation available to a code generation tool
contains adversarial context.  Weighting baseline and adversarial VGR
by their respective proportions yields an expected overall
vulnerability rate of:
\begin{equation*}
  \text{Expected VGR} = 0.05 \times 37.4\% + 0.95 \times
  3.5\% = 5.2\%
\end{equation*}
This represents a 49\% relative increase over the 3.5\% baseline, a
meaningful elevation in risk given the scale of AI-assisted
development adoption.

\subsection{Organizational Practices}
Technical defenses require complementary process controls:
mandatory security review of AI-generated code in high-risk functions,
context auditing treating third-party documentation with the same
scrutiny as third-party code, trust boundaries treating AI output as
untrusted input, and developer training on adversarial pattern
recognition.

\subsection{Limitations}
In terms of limitations, the experiments used Python and JavaScript
only and may not generalize to other languages. The models evaluated
represent versions available as of December 2024, and newer iterations
may exhibit different vulnerability profiles. Prompt contexts were
limited to 512 tokens, so the positional effects observed here may
differ in longer production contexts (4K--32K tokens). Finally, the
defense framework reflects detection capabilities at a fixed point in
time; adaptive adversaries may develop evasion strategies, and
longitudinal evaluation will be needed to maintain effectiveness.

\section{Related Work}
\label{sec:related}

\subsection{Code Generation Security}
Pearce et al.~\cite{pearce2022} documented systematic security issues
in GitHub Copilot, finding 40\% baseline vulnerability rates across 89
security-relevant scenarios.  Their work focused on prompt engineering
without investigating contextual manipulation or transferability.
Cotroneo et al.~\cite{cotroneo2024} demonstrated training-time
poisoning achieving 41\% VGR with only 3\% corrupted training data,
but these require infrastructure access not needed by our
inference-time approach.

\subsection{Prompt Injection Attacks}
Greshake et al.~\cite{greshake2023} defined indirect prompt injection
as adversarial content that enters an LLM-integrated application
through retrieved external data rather than direct user input which is
the closest analog to our attack model.  Liu et al.~\cite{liu2024}
formalized a framework for evaluating prompt injection attacks and
defenses across multiple LLMs.  These works focus on general LLM
applications rather than code generation specifics, which our work
directly addresses.

\subsection{Adversarial Machine Learning}
Goodfellow et al.~\cite{goodfellow2015} established the theoretical
basis for adversarial examples; Carlini and Wagner~\cite{carlini2017}
demonstrated cross-model transfer.  Jia and Liang~\cite{jia2017}
introduced adversarial evaluation for reading comprehension using
distractor sentences generated without gradient-based access.  Our
work contributes transferability analysis for code generation under
contextual manipulation.

\subsection{Code Model Internals}

Paltenghi and Pradel~\cite{paltenghi2021} compared human and neural
attention in code models, showing that model attention differs
substantially from developer attention.  Schuster et
al.~\cite{schuster2021} demonstrated poisoning in code completion; our
positional analysis explains why pre-function placement is most
effective.

\section{Ethical Considerations}
\label{sec:ethical}

All experiments in this study were conducted exclusively using each
model's API (OpenAI, Hugging Face) using automated
researcher-controlled prompts scripts, without targeting production
systems, real user data, or third-party infrastructure.  No
vulnerabilities were deployed or tested outside of the isolated
experimental environment.

Attack techniques are disclosed at a level sufficient for defenders to
build detection and mitigation systems; detailed exploitation
templates are withheld from the public version of this work.
Vulnerability patterns identified will be reported to the relevant
model providers (OpenAI, Meta) under a 90-day coordinated disclosure
window prior to public dataset release.

Defense code and the labeled experimental dataset will be released as
open artifacts under the Apache~2.0 license to support community
replication and further research.  This approach balances scientific
transparency with responsible security disclosure.

\section{Conclusion}
\label{sec:conclusion}

This research establishes context-based adversarial manipulation as a
critical, exploitable vulnerability in AI code generators through
2,800 controlled experiments with rigorous statistical validation.
Key contributions include: (1)~$10.7\times$ adversarial VGR increase
over baseline; (2)~high cross-model transferability across both
open-source and commercial systems; (3)~identification of high-risk
context positions and linguistic patterns; (4)~a real-time detection
framework achieving 89.1\% detection at 0.3\% FPR.

We urge AI code generation providers to: (1)~implement contextual
analysis before code generation; (2)~provide security-focused model
variants with hardened training; (3)~enable user-configurable security
policies; (4)~participate in coordinated vulnerability disclosure
programs.  The research community must prioritize inference-time
security alongside traditional training-time robustness.

\section*{Acknowledgment}

The authors thank the Beacom College of Computer \& Cyber Sciences at
Dakota State University for computational resources supporting this
research.  Claude Sonnet-4.6 (Anthropic) was used for spelling,
grammar, and LaTeX formatting assistance; all research design,
experimental execution, data analysis, and conclusions are solely the
authors' work.

\balance

\printbibliography
\end{document}